\documentclass[aps,prd,groupedaddress,showpacs,graphicx,nofootinbib]{revtex4}
\usepackage{amssymb,graphics,graphicx,color}

\begin{document}

\title{Delta-N Formalism for Curvaton with Modulated Decay}

\author{Kazunori Kohri}
\affiliation{Cosmophysics group, Theory Center, IPNS, KEK,  
and The Graduate University for Advanced Study (Sokendai), Tsukuba 305-0801, Japan}
\author{Chia-Min Lin}
\affiliation{Department of Physics, Kobe University, Kobe 657-8501, Japan}
\author{Tomohiro Matsuda}
\affiliation{Laboratory of Physics, Saitama Institute of Technology, Fukaya, Saitama 369-0293, Japan}

\begin{abstract}
In this paper, the curvature perturbation generated by the modulated curvaton decay is
studied by a direct application of $\delta N$-formalism.
Our method has a sharp contrast with the {\it non-linear formalism} which may be regarded as an indirect usage of $\delta N$-formalism.
We first show that our method can readily reproduce results in previous works of
modulation of curvaton.
Then we move on to calculate the case where the curvaton mass 
(and hence also the decay rate) is modulated. 
The method can be applied to the calculation of the modulation in the
freezeout model, in which the heavy species are considered instead of
 the curvaton.
Our method explains curvaton and various modulation on an equal footing.
\end{abstract}
\maketitle

\section{Introduction}
It is widely believed that a stage of cosmic
 inflation~\cite{Lyth:2009zz} which happens in the very early universe is
necessary to explain the primordial curvature perturbation. 
During inflation the quantum fluctuations of light scalar fields are
expanded to become superhorizon classical fluctuations. 
One or more of those field fluctuations are supposed to be responsible
for the curvature perturbation; the field may not be the inflaton field
 which drives inflation, but the curvaton field $\sigma$
 \cite{Lyth:2001nq, Enqvist:2001zp, Moroi:2001ct} or the modulation
 field $\chi$ \cite{Dvali:2003em, Kofman:2003nx, Dvali:2003ar} which
 modulates the decay rate of the inflaton field.  

According to the $\delta N$-formalism \cite{Sasaki:1995aw, Sasaki:1998ug, Lyth:2004gb, Lyth:2005fi} (see also
\cite{Kodama:1997qw}), the curvature perturbation resulted from a field $\phi$ is given by
\begin{equation}
\zeta=\delta N = N_\phi \delta \phi + \frac{1}{2}N_{\phi\phi}(\delta \phi)^2+\frac{1}{6}N_{\phi\phi\phi}(\delta \phi)^3+\cdots,
\label{deltaN}
\end{equation}
where $N=\int d \ln a$ is the number of e-folds and $a$ denotes the
scale factor.
The spectrum of the perturbation is ${\cal P}_{\delta \phi}^{1/2} \sim
H/2\pi$ at the horizon exit. 
Here $\phi$ may be the conventional inflaton field ($\varphi$), 
the curvaton field ($\sigma$) or the modulation field ($\chi$) which are
light (compared with the Hubble parameter) during inflation.
The fluctuations of those fields can (eventually) affect $N$. 
The subscript means derivative with respect to
$\phi$. Note that $\delta \phi$ is calculated on a flat slice (gauge)
and $\zeta$ is defined to be the curvature perturbation on uniform
energy density slice. It is convenient to estimate the magnitude of the 
higher order effects by using the nonlinear parameters $f_{NL}$ and
$g_{NL}$ defined by  
\begin{equation}
\zeta=\zeta_g + \frac{3}{5}f_{NL} \zeta^2_g+\frac{9}{25}g_{NL}\zeta^3_g + \cdots,
\end{equation}
where $\zeta_g=N_\phi \delta \phi$ denotes the Gaussian (i.e, the first-order expansion
with respect to the Gaussian perturbation $\delta \phi$) part of $\zeta$. 
From Eq.~(\ref{deltaN}) we can see that\footnote{It is also possible to have more than one field ($\chi$ and $\sigma$, for example) which can affect $N$. In this case, besides $N_{\sigma\sigma}$ and $N_{\chi\chi}$ we woud also have $N_{\chi\sigma}$ and the corresponding $(6/5)f^{\sigma\chi}_{NL} \equiv N_{\chi\sigma}/(N_\chi N_\sigma)$.} 
\begin{eqnarray}
f_{NL}&=&\frac{5}{6}\frac{N_{\phi\phi}}{(N_\phi)^2}, \\
g_{NL}&=&\frac{25}{54}\frac{N_{\phi\phi\phi}}{(N_\phi)^3}.
\end{eqnarray}
Current experimental data gives (very) roughly \cite{Bennett:2012fp, Smidt:2010sv}
\begin{eqnarray}
|f_{NL}| &\lesssim& 100, \\
|g_{NL}| &\lesssim& 10^6.
\end{eqnarray}
In the near future, the PLANCK satellite is expected to reduce the bound to
$|f_{NL}| \lesssim 10$ and $|g_{NL}|  \lesssim 10^5$ \cite{Smidt:2010ra} if non-gaussianity is not detected\footnote{Note added: PLANCK satellite has recently released their data \cite{Ade:2013ydc, Ade:2013uln}. This has interesting implications for curvaton model in general. A curvaton with modulated decay width or mass may help to relax constraints imposed on curvaton scenario by PLANCK data (see \cite{Huang:2013yla} as an example).}.

In some recent works the question of combining curvaton and
inhomogeneous reheating scenarios has been studied in the light of 
the modulated curvaton decay \cite{Enomoto:2012uy, Langlois:2013dh,
Assadullahi:2013ey, Enomoto:2013qf}, where the non-linear formalism of the
component perturbations based on \cite{Lyth:2004gb}, has been used.

In this paper, we propose a direct method which is conceptually simple and straightforward.
In section \ref{section2}, we show that the method can be used to
produce previous results of \cite{Langlois:2013dh,
Assadullahi:2013ey, Enomoto:2013qf}, in which non-linear formalism has been used.
In section \ref{section3}, in addition to the modulated decay rate, we
consider the modulation caused by the modulated curvaton mass. 
In appendix \ref{a}, we compare the calculation in section
\ref{section2} and section \ref{section3}.
In appendix \ref{b}, we reproduce standard formulas for the conventional
curvaton. 
In section \ref{section4}, the modulation in the freezeout
model \cite{Dvali:2003ar} is solved when the massive species might not
dominate the density.
Our method explains the curvaton model and the various modulation
scenarios on an equal footing.

\section{Modulated decay rate of the curvaton model : unmodulated mass}
\label{section2}

In this section, we will calculate the curvature perturbation 
when the curvaton decay is modulated. 
We consider the case in which the decay rate is modulated by a light
scalar field $\chi$ through a coupling. 
The case of the modulated decay rate through a mass is considered in
the next section.

\begin{figure}[t]
\centering
\includegraphics[width=0.5\columnwidth]{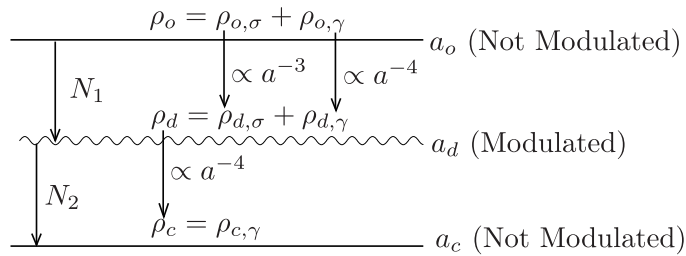}
 \caption{The modulated decay in the curvaton model is illustrated.
The curvaton density at $a_o$ is $\rho_{o,\sigma}=\frac{1}{2}m^2
 \sigma^2\ll \rho_{o,\gamma}$, which gives the curvaton density at the decay
$\rho_{d,\sigma}=\rho_{o,\sigma}X^{-3}$ where $X \equiv  (a_d/a_o)$.}
\label{fig:fig1}
\end{figure}

Consider a curvaton field $\sigma$ with a quadratic potential $V(\sigma)=(1/2)m^2 \sigma^2$.
The number of e-folds between curvaton oscillation $a_o$ to a uniform energy density time slice after curvaton decay at $a_c$, is given by
\begin{equation}
N = \ln \left( \frac{a_c}{a_o} \right) = \ln \left( \frac{a_d}{a_o} \right) + \ln \left( \frac{a_c}{a_d} \right) \equiv N_1+N_2,  
\label{eq1}
\end{equation}
where $a_d$ represents curvaton decay at $H \sim \Gamma$ and the two parts
are denoted by $N_1$ and $N_2$, respectively. After curvaton decay the universe is dominated by radiation therefore $H \propto a^{-2}$ and we can write $N_2$ as 
\begin{equation}
N_2 = \frac{1}{2} \ln \left( \frac{\Gamma}{H(a_c)} \right),        
\label{eq02}
\end{equation}
where we have used $H(a_d)=\Gamma$ and $H(a_c)$ corresponds to the uniform energy density slice (which does not depend on $\sigma$ or $\chi$). See also Fig.\ref{fig:fig1}.

By definition of the oscillating curvaton, when the curvaton start to
oscillate some time after inflation when $H \sim m$ at $a_o$, the energy
density is dominated by radiation $\gamma$. 
The energy densities are hence given by $\rho_{o} \sim \rho_{o,\gamma}=3m^2
M_p^2$ and $\rho_{o,\sigma}=(1/2)m^2\sigma^2$. 
Here the reduced Planck mass is $M_p=2.4 \times 10^{18}\mbox{ GeV}$
and $\sigma$ denotes the field value of the curvaton when it starts to
oscillate\footnote{This is often called $g(\sigma_\ast)$ where $\sigma_\ast$ denotes the field value of $\sigma$ at horizon exit. In this paper, we only consider the curvaton with a quadratic potential, therefore $g^\prime(\sigma_\ast) \equiv \partial g / \partial \sigma_\ast$ is a constant which is close to one when slow-roll condition is satisfied. In addition, one more derivative makes $g^{\prime\prime}(\sigma_\ast)=0$. Therefore we only consider $\sigma \sim \sigma_\ast$. The including of $g$ and its derivatives only makes our formulas look unnecessarily complicated withour gaining much.}. 
On the other hand, at curvaton decay ($H \sim \Gamma$) the energy
densities are given by $\rho_{d} = 3 \Gamma^2M_p^2$ and $\rho_{d,\sigma}
= (1/2)m^2\sigma(t_o)^2(a_d/a_o)^{-3}$ because the curvaton behaves like
cold matter during oscillation.

If we focus on radiation (which dilutes as $a^{-4}$) during this period
and define $X \equiv  (a_d/a_o)$ for notational simplicity, we can write 
\begin{equation}
X^4 \equiv \frac{\rho_{o,\gamma}}{\rho_{d,\gamma}}=\frac{3m^2M_p^2}{3\Gamma^2M_p^2-\frac{m^2\sigma^2}{2X^3}},    
\label{eq3}
\end{equation}
or 
\begin{equation}
3\Gamma^2 M_p^2X^4-\frac{1}{2}m^2\sigma^2 X=3m^2M_p^2.      
\label{eq4}
\end{equation}
For later usage, we define a quantity $r$ used widely in literatures of curvaton by 
\begin{equation}
r \equiv
 \frac{3\rho_{d,\sigma}}{4\rho_{d,\gamma}+3\rho_{d,\sigma}}
=\frac{3\left(\frac{1}{2}m^2\sigma^2  \right)X^{-3}}
{4\left[3\Gamma^2M_p^2-\left(\frac{1}{2}m^2\sigma^2\right)X^{-3}\right]
+3\left(\frac{1}{2}m^2\sigma^2\right)
X^{-3}}=\frac{3(\frac{1}{2}m^2\sigma^2)}{12\Gamma^2M_p^2 X^3-\frac{1}{2}m^2\sigma^2}. 
\label{eq5}
\end{equation}
In the oscillating curvaton model, $r$ is comparable to the ratio of the
curvaton energy density to the total energy density of the Universe at
the curvaton decay. 

The modulated curvaton decay is introduced by the function
$\Gamma(\chi)$, which becomes inhomogeneous in space due to the
modulation caused by a light field $\chi$.  
From Eq.~(\ref{deltaN}), the curvature perturbation to linear order is
given by $\zeta=N_\chi \delta \chi+ ...$, where other sources (e.g,
inflaton perturbation $\delta \varphi$ 
and the conventional curvaton perturbation $\delta \sigma$)
 are included in ``...''.
The last expression in Eq.~(\ref{eq5}) is convenient because it allows
us to calculate the derivatives of r with respect to the fields.
A very simple but handy relation which will be used frequently in this
paper is
\begin{equation}
\frac{12\Gamma^2M_p^2 X^3}{12\Gamma^2 M_p^2X^3 -\frac{1}{2}m^2\sigma^2}
=\frac{12\Gamma^2 M_p^2X^3 -\frac{1}{2}m^2\sigma^2
+\frac{1}{2}m^2\sigma^2}{12\Gamma^2M_p^2 X^3 -\frac{1}{2}m^2\sigma^2}
=1+\frac{r}{3}.
\label{handy}
\end{equation}

By making derivative of both sides of Eq.~(\ref{eq4}) with respect to $\chi$, we obtain 
\begin{equation}
6\Gamma \Gamma_\chi M_p^2X^4+12\Gamma^2M_p^2X^3 X_\chi -\frac{1}{2}m^2 \sigma^2 X_\chi=0,
\end{equation}
which implies
\begin{eqnarray}
N_{1\chi}&\equiv&\frac{\partial N_1}{\partial \chi}=(\ln X)_\chi=
 \frac{X_\chi}{X}  \nonumber\\
&=& \frac{6\Gamma^2 M_p^2 X^3 }{\frac{1}{2}m^2\sigma^2-12\Gamma^2 M_p^2X^3}
\frac{\Gamma_\chi}{\Gamma}\nonumber\\
&=&
-\left(\frac{1}{2}+\frac{r}{6}\right)\frac{\Gamma_\chi}{\Gamma},
\label{eq8}
\end{eqnarray}
where Eq.~(\ref{handy}) has been used.

$N_{2\chi}$ is evaluated by making a derivate of Eq.~(\ref{eq02}):
\begin{equation}
N_{2\chi}\equiv 
\frac{\partial N_2}{\partial \chi}=\frac{1}{2}\frac{\Gamma_\chi}{\Gamma}.
\label{n2}
\end{equation}
Therefore we have 
\begin{equation}
N_\chi=N_{1\chi}+N_{2\chi}=-\frac{1}{6}r\frac{\Gamma_\chi}{\Gamma}.
\label{eq9}
\end{equation}
This gives for example, Eq.~(31) in \cite{Enomoto:2013qf} and also consistent with \cite{Langlois:2013dh} and \cite{Assadullahi:2013ey}.

It is also straightforward to calculate the non-linear parameters. 
By making
derivative of Eq.~(\ref{eq9}) once more we have
\begin{equation}
N_{\chi\chi}
=\frac{r_\chi}{r}N_\chi
-\frac{1}{6}r \left(
	       \frac{\Gamma_{\chi\chi}\Gamma-\Gamma_\chi^2}{\Gamma^2}
	      \right).
\label{eq10}
\end{equation}
In order to evaluate $r_\chi\equiv \partial r/\partial \chi$, we make derivative of $r$ by using the last expression in Eq.~(\ref{eq5}) and keep in mind that $\Gamma(\chi)$ and $X(\chi)$ are now functions of $\chi$. We have
\begin{eqnarray}
r_\chi &=& \frac{-3\left( \frac{1}{2}m^2 \sigma^2 \right)}{12\Gamma^2 X^3 -\frac{1}{2}m^2\sigma^2}\frac{24\Gamma \Gamma_\chi X^3+36\Gamma^2 X^2 X_\chi}{12\Gamma^2X^3-\frac{1}{2}m^2 \sigma^2}  \nonumber\\
&=&-r \left[ 2\left(1+\frac{r}{3}\right)
\frac{\Gamma_\chi}{\Gamma}+3\left(1+\frac{r}{3}\right)\frac{X_\chi}{X} \right],
\end{eqnarray}
where we have used Eq.~(\ref{handy}) to obtain the second equality.
From Eq.~(\ref{eq9}) and (\ref{eq8}) we find useful relations
\begin{eqnarray}
\frac{\Gamma_\chi}{\Gamma}&=&-\frac{6}{r}N_\chi\nonumber\\
\frac{X_\chi}{X}&=&\left(\frac{3}{r}+1\right)N_\chi.
\end{eqnarray}
By using these relations we obtain
\begin{equation}
\label{r}
\frac{r_\chi}{r}=\frac{3-2r-r^2}{r}N_\chi.
\end{equation}
Therefore we find
\begin{equation}
\frac{6}{5}f_{NL}=\frac{N_{\chi\chi}}{N_\chi^2}
=\frac{3}{r}\left( 3-2\frac{\Gamma\Gamma_{\chi\chi}}{\Gamma_\chi^2}
	    \right)-2-r
=\left[\frac{9}{r}-2-r\right]
-\frac{6}{r}\left(\frac{\Gamma\Gamma_{\chi\chi}}{\Gamma_\chi^2}\right).
\end{equation}
By taking derivative of Eq.~(\ref{a3}) with respect to $\chi$, we have
\begin{equation}
N_{\chi\sigma}=\frac{2}{3}\frac{r_\chi}{\sigma}=-\frac{r(1-r)(3+r)\Gamma_\chi}{9\Gamma \sigma}, 
\end{equation}
where Eq.~(\ref{r}) has been used to evaluate $r_\chi$. Therefore by using Eqs.~(\ref{eq9}) and (\ref{a3}) we can obtain
\begin{equation}
\frac{6}{5}f^{\sigma\chi}_{NL} \equiv \frac{N_{\chi\sigma}}{N_\chi N_\sigma}=\frac{(1-r)(3+r)}{r}.
\label{crossng}
\end{equation}

We can carry on to calculate $g_{NL}$. Firstly we substitute Eq.~(\ref{r}) into Eq.~(\ref{eq10}) to write
\begin{equation}
N_{\chi\chi}
=\frac{3-2r-r^2}{r}N_\chi^2
-\frac{1}{6}r \left(
	       \frac{\Gamma_{\chi\chi}\Gamma-\Gamma_\chi^2}{\Gamma^2}
	      \right),
\end{equation} 
and making derivative with respect to $\chi$ once more. Then eliminate $r_\chi$ again by Eq.~(\ref{r}), finally we obtain
\begin{eqnarray}
\frac{54}{25}g_{NL}&=&\frac{N_{\chi\chi\chi}}{N_\chi^3}\nonumber\\
&=&\frac{1}{r^2}\left[135-54r-22r^2+10r^3+3r^4-18\left(9-2r-r^2\right)\frac{\Gamma \Gamma_{\chi\chi}}{\Gamma_\chi^2}+36\frac{\Gamma_{\chi\chi\chi}\Gamma^2}{\Gamma_\chi^3} \right].
\label{gnl1}
\end{eqnarray}
This gives Eq.~(43) in \cite{Assadullahi:2013ey} and Eq.~(28) in \cite{Langlois:2013dh}. It should be clear that our method is capable to reproduce previous results.



\section{Modulated decay rate of the curvaton model : modulated mass}
\label{section3}

\begin{figure}[t]
\centering
\includegraphics[width=0.5\columnwidth]{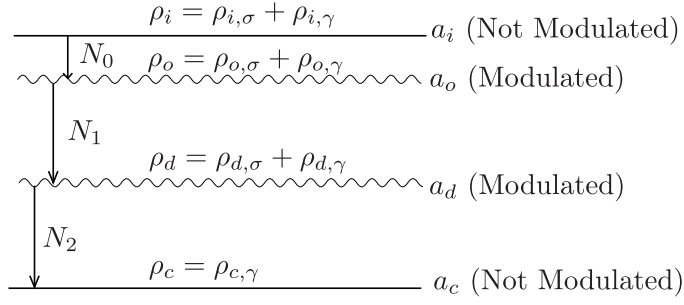}
 \caption{The modulated decay in the curvaton model is illustrated when
 the curvaton mass is modulated.}
\label{fig:fig2}
\end{figure}

In this section, we apply our method to the case where the modulated
decay rate is sourced by the modulated curvaton mass. This is more complicated
than the previous case because now the time slice when the curvaton
start to oscillate is also modulated.
{\it In that way, the quantities defined at $a_o$ may also depend on
$\chi$.}
We thus need to introduce $a_i$ before $a_o$, since not only $a_d$ but also
$a_o$ is modulated.
See also Fig.\ref{fig:fig2}

Before the curvaton oscillation, the curvaton is slow-rolling and we
assume that the curvaton density is scaling like $\rho_\sigma \propto a^{-\epsilon_w}$
just before the oscillation. 
In general $\epsilon_w$ is time-dependent in the radiation-dominated
Universe; however for the modulation at $a_0$ 
we only need $\epsilon_w$ defined very close to $a_o$.
Therefore in our calculation we can assume that $\epsilon_w$ is approximately a constant.

The e-folding number 
is now given by 
\begin{equation}
N = \ln \left( \frac{a_c}{a_i} \right) = \ln \left( \frac{a_o}{a_i}
					     \right)+
\ln \left( \frac{a_d}{a_o} \right) + \ln \left( \frac{a_c}{a_d} \right)
=N_0+N_1+N_2,
\end{equation}
where $N_2$ is again given by Eq.~(\ref{eq02}) and $N_{2\chi}$ is given by Eq.~(\ref{n2}).

In order to calculate $N_{0\chi}$,
we set the fraction of the energy density at $a_i$ as
\begin{equation}
f_{i}\equiv \frac{\rho_{i,\sigma}}{\rho_i}
=\frac{\rho_{i,\sigma}}{3H^2_i M_p^2}.
\end{equation}
If we define 
\begin{equation}
N_0=\ln \left(\frac{a_o}{a_i}\right) \equiv \ln X_0,
\end{equation}
we find
\begin{equation}
\label{x0-ini}
X_0^4=\frac{\rho_{i,\gamma}}{\rho_{o,\gamma}}
=\frac{3H_i^2(1-f_{i})}{3m^2(1-f_{o})},
\end{equation}
where
\begin{eqnarray}
f_{o}&\equiv&\frac{\rho_{o,\sigma}}{3m^2 M_p^2}=
\frac{\rho_{i,\sigma}X_0^{-\epsilon_w}}{3m^2 M_p^2} \nonumber\\
&&\frac{\partial f_{o}}{\partial \chi}
=-\left(\epsilon_w\frac{X_{0\chi}}{X_0}\right)f_{o}\nonumber\\
f_{i}&\equiv& \frac{\rho_{i,\sigma}}{\rho_i}
=\frac{\rho_{i,\sigma}}{3H^2_i M_p^2}\nonumber\\
&&\frac{\partial f_{i}}{\partial \chi}=
2\frac{m_\chi}{m}f_{i}.
\end{eqnarray}
Here, for the curvaton density we assume\footnote{This is in sharp
contrast to the ``freezeout and mass-domination 
model'' in \cite{Dvali:2003ar}, which uses $\rho_\sigma\propto m^4$ for
the initial condition. 
 We will compare those models in the next section to show that
 our method explains those different scenarios on an equal footing.}
$\rho_{i,\sigma}\propto m^2$.
We can find from Eq.(\ref{x0-ini}):
\begin{equation}
4X_0^3 X_{0\chi}=
-\frac{\partial f_i}{\partial \chi}\frac{\rho_{i}}{\rho_{o,\gamma}}
-2\frac{m_\chi}{m}\frac{\rho_{i,\gamma}}{\rho_{o,\gamma}}
+\frac{\partial f_o}{\partial \chi}\frac{\rho_{i,\gamma}\rho_{o}}{\rho_{o,\gamma}^2}.
\end{equation}
Considering $\frac{\rho_{i,\gamma}}{\rho_{o,\gamma}} =X_0^{4}$, we find
\begin{eqnarray}
4\frac{X_{0\chi}}{X_0}&=&
-2\frac{m_\chi}{m}\frac{(\rho_{i}f_i)}{\rho_{i,\gamma}}
-2\frac{m_\chi}{m}
-\epsilon_w\frac{X_{0\chi}}{X_0}\frac{(\rho_{o}f_o)}{\rho_{o,\gamma}}.
\end{eqnarray}
Solving that equation for $\frac{X_{0\chi}}{X_0}$,  we obtain
\begin{eqnarray}
\frac{X_{0\chi}}{X_0}&=&
-\frac{1}{2}(1-r_o)\frac{m_\chi}{m}\left(1+\frac{\rho_{i,\sigma}}{\rho_{i,\gamma}}\right),
\end{eqnarray}
where 
\begin{eqnarray}
r_o&\equiv& \frac{\epsilon_w\rho_{o,\sigma}}{4\rho_{o,\gamma}+\epsilon_w
 \rho_{o,\sigma}}\nonumber\\
(1-r_o)&\equiv& \frac{4\rho_{o,\gamma}}{4\rho_{o,\gamma}+\epsilon_w
 \rho_{o,\sigma}}.
\end{eqnarray}
Finally, we have
\begin{eqnarray}
N_{0\chi}&=&(\ln X_0)_\chi\nonumber\\
&=&-\frac{1}{2}(1-r_o)\frac{m_\chi}{m}
\left(1+\frac{\rho_{i,\sigma}}{\rho_{i,\gamma}}\right). 
\end{eqnarray}
For the usual set-ups $f_i\simeq f_o \simeq 0$, the result shows 
\begin{equation}
N_{0\chi}\simeq -\frac{1}{2} \frac{m_\chi}{m}.
\end{equation}

For the second stage (i.e, during the curvaton oscillation), we define 
\begin{equation}
N_1=\ln \left(\frac{a_d}{a_o}\right)=\ln X_1.
\end{equation}
Then 
\begin{equation}
\label{x1-ini}
X_1^4=\frac{\rho_{o,\gamma}}{\rho_{d,\gamma}}
=\frac{3m^2(1-f_{o})}{3\Gamma^2(1-f_{d})},
\end{equation}
where
\begin{eqnarray}
f_{d}&\equiv& \frac{\rho_{d,\sigma}}{\rho_{d}}=
\frac{\rho_{i,\sigma}X_1^{-3}X_o^{-\epsilon_w}}{3\Gamma^2 M_p^2}\nonumber\\
&&\frac{\partial f_{d}}{\partial \chi}
=\left(-3\frac{X_{1\chi}}{X_1}-\epsilon_w\frac{X_{0\chi}}{X_0}
+2\frac{m_\chi}{m}
-2\frac{\Gamma_\chi}{\Gamma}\right)f_{d}\nonumber\\ 
f_{o}&\equiv& \frac{\rho_{o,\sigma}}{\rho_{o}}=
\frac{\rho_{i,\sigma}X_0^{-\epsilon_w}}{3m^2 M_p^2}\nonumber\\
&&\frac{\partial f_{o}}{\partial \chi}
= \left(-\epsilon_w\frac{X_{0\chi}}{X_0}\right)f_o.
\end{eqnarray}
We thus find from Eq.(\ref{x1-ini}):
\begin{equation}
4X_1^3 X_{1\chi}=\left(2\frac{m_\chi}{m}-2\frac{\Gamma_\chi}{\Gamma}\right)
\frac{\rho_{o,\gamma}}{\rho_{d,\gamma}}
-\frac{\partial f_o}{\partial \chi}\frac{\rho_{o}}{\rho_{d,\gamma}}
+\frac{\partial f_d}{\partial \chi}\frac{\rho_{o,\gamma}\rho_{d}}{\rho_{d,\gamma}^2}.
\end{equation}
By using $\frac{\rho_{o,\gamma}}{\rho_{d,\gamma}} =X_1^4$, we find
\begin{eqnarray}
4\frac{X_{1\chi}}{X_1}&=&
\left(2\frac{m_\chi}{m}-2\frac{\Gamma_\chi}{\Gamma}\right)
-\epsilon_w\frac{X_{o\chi}}{X_0}
\frac{(\rho_{o}f_o)}{\rho_{o,\gamma}}
+\left(-3\frac{X_{1\chi}}{X_1}-\epsilon_w\frac{X_{0\chi}}{X_0}
+2\frac{m_\chi}{m}-2\frac{\Gamma_\chi}{\Gamma}\right)
\frac{(\rho_d f_d)}{\rho_{d,\gamma}}.
\end{eqnarray}
Although not mandatory, we are going to assume $\epsilon_w\ll 1$ in this paper.
Solving the equation for $\frac{X_{1\chi}}{X_1}$,  we find
\begin{eqnarray}
\frac{X_{1\chi}}{X_1}&\simeq&-
\frac{2(\rho_{d,\gamma}+\rho_{d,\sigma})}{4\rho_{d,\gamma}+3\rho_{d,\sigma}}\
\left(\frac{\Gamma_\chi}{\Gamma}
-\frac{m_\chi}{m}\right)\nonumber\\
&=&-\left(\frac{1}{6}r+\frac{1}{2}\right)
\left(\frac{\Gamma_\chi}{\Gamma}
-\frac{m_\chi}{m}\right).
\end{eqnarray}
where $r$ is given by Eq.~(\ref{eq5}) and we have used the relation
\begin{equation}
\frac{1}{6}r+\frac{1}{2}= \frac{2\rho_{d,\gamma}+2\rho_{d,\sigma}}
{4\rho_{d,\gamma}+3\rho_{d,\sigma}}=\frac{2\rho_{d}}
{4\rho_{d,\gamma}+3\rho_{d,\sigma}}.
\end{equation}
Therefore
\begin{eqnarray}
N_{1\chi}&=&(\ln X_1)_\chi\nonumber\\
&\simeq&-\left(\frac{1}{6}r+\frac{1}{2}\right)
\left(\frac{\Gamma_\chi}{\Gamma}
-\frac{m_\chi}{m}\right).
\end{eqnarray}
Our final result is 
\begin{eqnarray}
N_\chi&=&N_{0\chi}+N_{1\chi}+N_{2\chi}\\
&\simeq&
\frac{r}{6}\left[-\frac{\Gamma_\chi}{\Gamma}
+\frac{m_\chi}{m}\right]
\end{eqnarray}

To understand the result, imagine an explicit form of $\Gamma$.
Just for instance, one may assume $\Gamma\propto \lambda m^n$. This gives\footnote{Even if $\epsilon_w$ is not negligible, we can define $n'=(n+\epsilon_w)$ to
obtain a simple formula
\begin{eqnarray}
N_\chi
&\simeq&
\frac{1}{6}r\left(1-n'\right)
\frac{m_\chi}{m}.
\end{eqnarray}
}
\begin{eqnarray}
N_\chi
&\simeq&
\frac{1}{6}r\left(1-n
\right)\frac{m_\chi}{m}.
\end{eqnarray}
It might be interesting to note here that when $n=1$, the effects of modulated curvaton oscillation and decay cancel each other out.

For the non-linear parameter, we make derivative with respect to $\chi$ once more to obtain
\begin{equation}
N_{\chi\chi}=\frac{1}{6}r_\chi(1-n)\frac{m_\chi}{m}+\frac{1}{6}r(1-n)\left( \frac{m_{\chi\chi}m-m_\chi^2}{m^2} \right),
\end{equation}
where by using similar method as in the previous section, we can obtain
\begin{equation}
\frac{r_\chi}{r}=\frac{3-2r-r^2}{r}N_\chi.                     
\end{equation}
Interestingly, $r_\chi$ has the same form as what was given in Eq.~(\ref{r})\footnote{This implies $f^{\sigma \chi}_{NG}$ also has the same form as Eq.~(\ref{crossng}). However they are not really the same because the corresponding $N_\chi$ are different.}.
The nonlinear parameters are hence given by
\begin{equation}
\frac{6}{5}f_{NL}=-r-2+\frac{3}{r}-\frac{6}{r(1-n)}+\frac{6}{r(1-n)}\frac{m_{\chi\chi}m}{m^2_\chi}.
\end{equation}
and
\begin{eqnarray}\nonumber
\frac{54}{25}g_{NL}&=&\frac{N_{\chi\chi\chi}}{N_\chi^3}=3r^2+10r-4-\frac{18}{r}+\frac{18}{1-n}+\frac{36}{r(1-n)}+\frac{9}{r^2}-\frac{54}{r^2(1-n)}+\frac{72}{r^2(1-n)^2}  \\ 
&&+\left( -\frac{18}{1-n}-\frac{36}{r(1-n)}+\frac{54}{r^2(1-n)}-\frac{108}{r^2(1-n)^2} \right)\frac{mm_{\chi\chi}}{m_\chi^2}+ 36\frac{m_{\chi\chi\chi}m^2}{m_\chi^3}.
\end{eqnarray}

\section{Freezeout model with the modulated mass}
\label{section4}
The modulated decay scenario of the freezeout model
is explained by a massive particle species $\psi$, whose
mass $M$ and 
the decay rate $\Gamma$ may depend on $\chi$.
Initially the massive species $\psi$ are subdominant.

\begin{itemize}
\item In the original model~\cite{Dvali:2003ar} it has been assumed that
      $\psi$ were thermalized at some early time ($T>T_f>T_n$) and becomes
      non-relativistic at the temperature 
      $T_n\simeq M$~\cite{Dvali:2003ar}. Here $T_f$ is the
      freezeout temperature. Then the density of 
the massive species $\psi$ at that moment is 
\begin{equation}
\rho_{\psi}(T_n)=Mn_\psi(T_n)\simeq M^4,
\end{equation}
where $M$ denotes the mass of $\psi$ species.
We usually have the total density $\rho(T_n)>\rho_\psi(T_n)$.
Replacement from the modulated curvaton scenario is:
\begin{eqnarray}
\rho_{o,\sigma} \simeq \frac{1}{2}m^2\sigma^2 &\rightarrow& 
\rho_{n,\psi} \propto M^4,
\end{eqnarray}
which is defined at the beginning of the
      scaling $\rho_\psi\propto a^{-3}$.
\item Normally, one will assume that the freezeout occurs after $\psi$
      becomes non-relativistic ($M>T_f$).
In that case we find the Boltzmann suppression: 
\begin{equation}
n_\psi(T_f)= g\left(\frac{MT_f}{2\pi}\right)^{3/2}\exp\left(-\frac{M}{T_f}\right),
\end{equation}
where $g$ depends on the model.
In that case we find
\begin{equation}
\label{non-eq-massdm}
\rho_{\psi}(T_f)\simeq M
 n_\psi(T_f)=gM^{5/2}\left(\frac{T_f}{2\pi}\right)^{3/2}
\exp\left(-\frac{M}{T_f}\right).
\end{equation}
It is possible to introduce a factor~\cite{Kolb:1990vq, Jungman:1995df}
\begin{eqnarray}
x_f&\equiv& \frac{M}{T_f}\\
&\simeq& 25 +\ln\frac{M}{\mathrm{TeV}}+\ln\frac{\langle \sigma
 v \rangle}{\mathrm{(TeV)^{-2}}},
\end{eqnarray}
where $\langle \sigma v \rangle$ is the thermal-averaged annihilation
      cross section, 
      and write
\begin{equation}
\rho_\psi(T_f)= M^{4}\left[\frac{g}{\left(2\pi x_f\right)^{3/2}}e^{-x_f}\right].
\end{equation}
When the cross section does not depend on $M$, the replacement from
      the modulated curvaton scenario becomes 
\begin{eqnarray}
\rho_{o,\sigma} \simeq \frac{1}{2}m^2\sigma^2 &\rightarrow& 
\rho_{f,\psi} \propto M^3(x_f)^{-3/2},
\end{eqnarray}
where the trivial identity $e^{-\ln M}=M^{-1}$ has been used.
The modulation about the parameter $x_f$ is weak and negligible.
The ratio of the energy density at the freezeout temperature is
\begin{equation}
f_{f}\equiv \frac{\rho_{f,\psi}}{\rho_f}
\propto \frac{(x_f)^{5/2}}{M}.
\end{equation}
Alternatively, one may assume
$\langle\sigma v\rangle\propto M^{-2}$ and find
\begin{eqnarray}
\rho_{f,\psi} \propto M^5(x_f)^{-3/2},
\end{eqnarray}
which gives the ratio
\begin{equation}
f_{f}\equiv \frac{\rho_{f,\psi}}{\rho_f}
\propto (x_f)^{5/2} M.
\end{equation}
Note that $x_f=1$ reproduces the first scenario ($\rho_\psi\propto M^4$
      and $f\propto M^0$). 
In this paper we are assuming instant trandition between phases and the
      simple $M$-dependence for simplicity.
The actual calculation has to be highly model-dependent. 
\end{itemize}

In this section we mainly consider the first scenario (or equivalently
$x_f=1$ in the second scenario), since the original
paper about the freezeout model~\cite{Dvali:2003ar} considers that
possibility. 
For the freezeout model, we use the subscript ``F'' to denote the time
when the density starts to scale like $\rho_\psi\propto a^{-3}$.
If we define 
\begin{equation}
N_0=\ln \left(\frac{a_F}{a_i}\right) \equiv \ln X_0,
\end{equation}
we find
\begin{equation}
X_0^4= \frac{\rho_{i,\gamma}}{\rho_{F,\gamma}}
=\frac{\rho_{i,\gamma}}{\rho_{F}(1-f_F)},
\end{equation}
where $\rho_{i}$ is defined just before $\rho_F$ so that the density
scaling is well approximated by $\rho_\gamma\propto a^{-4}$.
We thus find for $\partial \rho_{i,\gamma}/\partial \chi\simeq 0$:
\begin{eqnarray}
4\frac{X_{0\chi}}{X_0}&=&
-\frac{1}{\rho_{F}} \frac{\partial \rho_{F}}{\partial \chi}
+\frac{1}{1-f_F}\frac{\partial f_F}{\partial
\chi}.
\end{eqnarray}
Therefore 
\begin{equation}
N_{0\chi}=(\ln X_0)_\chi=-\frac{1}{4\rho_F} \frac{\partial
 \rho_{F}}{\partial \chi}
+\frac{1}{4(1-f_F)}\frac{\partial f_F}{\partial \chi}.
\end{equation}
After the freezeout, we define
\begin{equation}
N_1=\ln \left(\frac{a_d}{a_F}\right)=\ln X_1.
\end{equation}
Then, we find
\begin{equation}
X_1^4
\simeq \frac{\rho_F(1-f_F)}{3\Gamma^2M_p^2(1-f_{d})}.
\end{equation}
We thus find 
\begin{equation}
4\frac{X_{1\chi}}{X_1}=
\frac{1}{\rho_F}\frac{\partial \rho_{F}}{\partial \chi}
-\frac{1}{1-f_F}\frac{\partial f_F}{\partial \chi}
-2\frac{\Gamma_\chi}{\Gamma}
+\frac{1}{(1-f_d)}\frac{\partial f_d}{\partial\chi}, 
\end{equation}
where 
\begin{eqnarray}
f_{d}&\equiv& \frac{\rho_{d,\psi}}{\rho_{d}}=
\frac{\rho_{F,\psi}X_1^{-3}}{3\Gamma^2 M_p^2}\\
&&\frac{\partial f_{d}}{\partial \chi}
=\left(-3\frac{X_{1\chi}}{X_1}+\frac{1}{\rho_{F,\psi}}
\frac{\partial \rho_{F,\psi}}{\partial \chi}
-2\frac{\Gamma_\chi}{\Gamma}\right)f_{d}.
\end{eqnarray}
We find\footnote{In the simplest case one may assume $f_F\propto M^k$ to find
$\partial f_F/\partial \chi =k \frac{M_\chi}{M}$.
The first scenario gives $k=0$, while the second scenario
suggests $k\ne 0$.}
\begin{eqnarray}
4\frac{X_{1\chi}}{X_1}&=&
\left(\frac{1}{\rho_F}\frac{\partial \rho_{F}}{\partial \chi}
-2\frac{\Gamma_\chi}{\Gamma}\right)
-\frac{1}{1-f_F}\frac{\partial f_F}{\partial \chi}\\ 
&&+\left(-3\frac{X_{1\chi}}{X_1}+
\frac{1}{\rho_{F,\psi}}\frac{\partial \rho_{F,\psi}}{\partial \chi}
-2\frac{\Gamma_\chi}{\Gamma}\right)
\frac{f_d}{(1-f_d)}.
\end{eqnarray}
Solving the equation for $\frac{X_{1\chi}}{X_1}$,  we find
\begin{eqnarray}
\frac{X_{1\chi}}{X_1}&=&-
\frac{2(\rho_{d,\gamma}+\rho_{d,\psi})}{4\rho_{d,\gamma}+3\rho_{d,\psi}}\
\frac{\Gamma_\chi}{\Gamma}
+\frac{r}{3\rho_{F,\psi}}\frac{\partial \rho_{F,\psi}}{\partial \chi}
+(1-r)\left(\frac{1}{4\rho_F}\frac{\partial \rho_{F}}{\partial \chi}\right)
-\frac{1-r}{4(1-f_F)}\frac{\partial f_F}{\partial \chi}\\
&=&-\left(\frac{1}{6}r+\frac{1}{2}\right)
\frac{\Gamma_\chi}{\Gamma}
+\frac{r}{3\rho_{F,\psi}}\frac{\partial \rho_{F,\psi}}{\partial \chi}
+(1-r)\left(\frac{1}{4\rho_F}\frac{\partial \rho_{F}}{\partial
       \chi}\right)
-\frac{1-r}{4(1-f_F)}\frac{\partial f_F}{\partial \chi}.
\end{eqnarray}
Therefore
\begin{eqnarray}
N_{1\chi}=(\ln X_1)_\chi
&=&-\left(\frac{1}{6}r+\frac{1}{2}\right)
\frac{\Gamma_\chi}{\Gamma}
+\frac{r}{3\rho_{F,\psi}}\frac{\partial \rho_{F,\psi}}{\partial \chi}
+(1-r)\left(\frac{1}{4\rho_F}\frac{\partial \rho_{F}}{\partial
       \chi}\right)
-\frac{1-r}{4(1-f_F)}\frac{\partial f_F}{\partial \chi}.
\end{eqnarray}
Our final result is 
\begin{eqnarray}
N_\chi&=&N_{0\chi}+N_{1\chi}+N_{2\chi}\\
&\simeq&
r\left[-\frac{1}{6} \frac{\Gamma_\chi}{\Gamma}
+\frac{\partial \rho_{F,\psi}/\partial \chi}{3\rho_{F,\psi}}
-\frac{\partial \rho_{F}/\partial \chi}{4\rho_{F}}
+\frac{1}{4}\frac{\partial f_F}{\partial \chi}
\right].
\end{eqnarray}

To understand the result, consider explicit forms of $\rho_{F,\psi}$.
\begin{itemize}
\item For $\rho_{F,\psi}\propto M^4$, $k=0$ and $\rho_F\propto M^4$, we
      find 
\begin{eqnarray}
N_\chi
&\simeq&
-\frac{1}{6}r\left(\frac{\Gamma_\chi}{\Gamma} -2\frac{M_\chi}{M}
\right),
\end{eqnarray}
where $r=1$ reproduces the original calculation of \cite{Dvali:2003ar}.
The result is consistent with the conventional
calculation of the mass-domination and the freezeout
      scenario~\cite{Dvali:2003ar}.

\item Replacing $\rho_{F,\psi}\rightarrow \rho_{o,\sigma}\propto m^2$ and 
$\rho_F\rightarrow \rho_o \propto m^2$ with $k=0$,\footnote{We find
      $f_F\sim \rho_{F,\sigma}/\rho_F\sim (m^2\sigma^2)/(m^2M_p^2)\propto m^0$.} it
      reproduces the modulated curvaton in the previous section:
\begin{eqnarray}
N_\chi&=&-\frac{1}{6}r \left(
\frac{\Gamma_\chi}{\Gamma}-\frac{m_\chi}{m}\right),
\end{eqnarray}
where the $\epsilon_w$-dependence does not appear in the above
      calculation.

\item 
In a more realistic calculation one must consider
Eq.~(\ref{non-eq-massdm}) and the cross section using the numerical 
      methods, which may shift the coefficients~\cite{Vernizzi:2003vs}.
\end{itemize}

In the multi-field modulation model we find
$\delta N = \sum_i N_{\chi_i}\delta \chi_i+\sum_i \sum_j
N_{\chi_i\chi_j} \delta \chi_i\delta \chi_j+...$;
\begin{eqnarray}
\label{multi-mod}
N_{\chi_i}&\simeq&
-\frac{1}{6}r \frac{\Gamma_{\chi_i}}{\Gamma}
+\frac{r}{3}\frac{1}{\rho_{F,\psi}}\frac{\partial
\rho_{F,\psi}}{\partial \chi_i}
-\frac{r}{4}\frac{1}{\rho_{F}}\frac{\partial
\rho_{F}}{\partial \chi_i}+\frac{r}{4}\frac{\partial f_F}{\partial \chi}
\\
&\simeq& 
r\left[-\frac{1}{6} \ln \Gamma
+\frac{1}{3}\ln \rho_{F,\psi}-\frac{1}{4}\ln \rho_{F}
+\frac{k}{4}\ln f_F
\right]_{\chi_i}
\\
N_{\chi_i\chi_j}&\simeq&
\frac{r_{\chi_j}}{r}N_{\chi_i}
+r\left[-\frac{1}{6} \ln \Gamma
+\frac{1}{3}\ln \rho_{F,\psi}-\frac{1}{4}\ln
\rho_{F} +\frac{k}{4}\ln f_F\right]_{\chi_i\chi_j}
\\
&\simeq&-\frac{(r-1)(r+3)}{r}N_{\chi_i}N_{\chi_j}
+r\left[-\frac{1}{6} \ln \Gamma
+\frac{1}{3}\ln \rho_{F,\psi}-\frac{1}{4}\ln \rho_{F}
+\frac{k}{4}\ln f_F\right]_{\chi_i\chi_j}.
\end{eqnarray}
In our direct method, it is very easy to evaluate the higher derivatives.

Notable application of the above result is the conventional curvaton
(i.e, the curvaton without extra modulation).
For the curvaton one needs just a simple replacement $\chi_j\rightarrow
\sigma$.
The curvaton hypothesis gives $\frac{\Gamma_{\sigma}}{\Gamma}=0$ and
$\frac{1}{\rho_F}\frac{\partial \rho_{F}}{\partial
       \chi_i}\simeq 0$.
Then one will find
\begin{eqnarray}
N_{\sigma}&\simeq&
\frac{r}{3}\frac{1}{\rho_{F,\sigma}}\frac{\partial
\rho_{F,\sigma}}{\partial \sigma}=\frac{2r}{3}\frac{\delta \sigma}{\sigma}.
\end{eqnarray}
In the above formalism {\it the curvaton mechanism is calculated as a
specific example of the modulation.}
In that way, {\it the mixed perturbations of the curvaton and the modulation
are calculated in our formalism as the multi-field modulation.}

\section{conclusion and discussion}
\label{conclusion}

In this paper, we proposed a direct application of $\delta N$
formalism that can be used to calculate the curvature perturbation from
the curvaton (or the heavy species) with modulation.
We calculated for the first time the case where
the modulated curvaton decay is due to the modulated curvaton mass
and obtained non-linear parameters.
Our method can be compared with the calculation based on the
non-linear formalism of the component perturbations.

Although we consider a quadratic potential for the
curvaton which dilutes like cold matter when oscillate,
in principle the method can be extended to more general cases once we
specify the modulation and the dilution 
behavior~\cite{Enomoto:2013qf, Matsuda:2009yt}.
Our method explains curvaton and various modulation models on an equal
footing and provides a convenient way to calculate the cosmological
perturbations in the multi-component Universe.

One of the natural cosmological expectations would be that a non-relativistic matter is created and its density starts dominating late after reheating. The original curvaton mechanism is based on that simple expectation; however the original curvaton mechanism requires significant isocurvature perturbation of the matter density. As a consequence, the "non-relativistic matter" is usually replaced by a "sinusoidal oscillation" whose amplitude must be inhomogeneous in space.

In the light of the cosmological model building, the original curvaton conjecture seems to be quite restrictive. There could be a deviation from the sinusoidal oscillation (i.e, the scaling of the density could be different due to a deviation from the quadratic potential) or the "non-relativistic matter density" could be the "conventional particle" that is created by the usual thermal process.

The former possibility (deviation from the matter scaling) has been discussed in \cite{Enomoto:2013qf}, and the latter (non-relativistic "particle" from the conventional thermal process) has been discussed in this paper. The method provided in this paper makes the calculation in \cite{Enomoto:2013qf} drastically easy. Obviously, the models discussed in this paper (and in \cite{Enomoto:2013qf}) are expanding to a great extent the application of the original curvaton mechanism.

Recently a significant extension of the curvaton scenario has been discussed
in \cite{Dimopoulos:2011gb}, in which an inflationary stage is considered
for the curvaton mechanism instead of the oscillation.
In the name of the ``curvaton'', the curvaton inflation converts
isocurvature perturbations that already exists at the beginning of the
secondary inflation into curvature perturbations whose wavelength is far
beyond the reach of the secondary (curvaton)
inflation.\footnote{Applications of the inflating curvaton can be found in
\cite{Furuuchi:2011wa, Dimopoulos:2012nj, Kohri:2012yw}.
In contrast to the conventional curvaton, the non-Gaussianity parameter
$f_{NL}$ is expected to be positive in the inflating curvaton, which
helps PBH generation in the curvaton mechanism~\cite{Kohri:2012yw}.} 
Higher order perturbations are calculated in \cite{Enomoto:2013qf},
although the calculation depends on the indirect method of the
non-linear formalism.
The direct calculation of the $\delta N$ formalism presented in this
paper may have the potential application to the inflating curvaton
mechanism, which can include any kind of modulation at the same time.

\section*{Acknowledgement}
K.K. is supported in part by Grantin- Aid for Scientific
research from the Ministry of Education, Science, Sports, and
Culture (MEXT), Japan, No. 21111006, No. 22244030, and No. 23540327.
CML would like to thank Chuo University for hospitality during the time
this work has been done.
T.M wishes to thank K.~Shima for encouragement, and his colleagues at
Lancaster university for their kind hospitality and many invaluable
discussions. 

\appendix

\section{More about the calculation details}

\subsection{Comparison between calculations of Section \ref{section2} and
  Section \ref{section3}}
\label{a}

We show that by using the same method, we can have a slightly different
way to obtain results given in section \ref{section2}. 
This calculation is closer to the one used in section \ref{section3}. 

At curvaton oscillating $a_o$, we set the fraction of the curvaton
energy density to be 
\begin{equation}
f_{\sigma,o}\equiv \frac{\rho_{o,\sigma}}{\rho_\sigma}
=\frac{\rho_{o,\sigma}}{3m^2 M_p^2}.
\end{equation}
If we define
\begin{equation}
N_1=\ln \left(\frac{a_d}{a_o}\right)=\ln X,
\end{equation}
we find
\begin{equation}
X^4=\frac{\rho_{o,\gamma}}{\rho_{d,\gamma}}
=\frac{3m^2M_p^2(1-f_{o})}{3\Gamma^2M_p^2(1-f_{d})},
\end{equation}
where
\begin{eqnarray}
f_{d}&=&\frac{\rho_{o,\sigma}}{3\Gamma^2 M_p^2}X^{3}\\
\frac{\partial f_{d}}{\partial \chi}
&=&\left(3\frac{X_\chi}{X}-2\frac{\Gamma_\chi}{\Gamma}\right)f_{d}\\
\frac{\partial f_{o}}{\partial \chi}&=& 0.
\end{eqnarray}
Therefore
\begin{eqnarray}
4\frac{X_\chi}{X}&=&-\frac{2\frac{\Gamma_\chi}{\Gamma}\rho_{d,\gamma}
-3\frac{X_\chi}{X}\rho_{d,\sigma}+2\frac{\Gamma_\chi}{\Gamma}
\rho_{d,\sigma}}{\rho_{d,\gamma}},
\end{eqnarray}
where $\rho_{d,\gamma}= 3m^2M_p^2(1-f_o)X^{-4}$.
We thus find 
\begin{eqnarray}
\frac{X_\chi}{X}&=&\frac{-2\frac{\Gamma_\chi}{\Gamma}\rho_{d}}
{4\rho_{d,\gamma}+3\rho_{d,\sigma}}=-\left(\frac{1}{6}r+\frac{1}{2}\right)
\frac{\Gamma_\chi}{\Gamma}.
\end{eqnarray}
Finally, we find
\begin{equation}
N_{1\chi}=(\ln X)_\chi =-\left(\frac{1}{6}r+\frac{1}{2}\right)
\frac{\Gamma_\chi}{\Gamma}.
\end{equation}
The remaining calculations are the same as those in section \ref{section2}.


\subsection{Conventional (oscillating) curvaton}

\label{b}

In this appendix, we rederive some familiar formulas of oscillating curvaton \cite{Sasaki:2006kq} by using our method.
Let us start from Eq.~(\ref{eq4}),
\begin{equation}
3\Gamma^2 M_p^2 X^4-\frac{1}{2}m^2\sigma^2 X=3m^2 M_p^2.   
\end{equation}
We make derivative to both sides with respect to $\sigma$ to obtain
\begin{equation}
12 \Gamma^2 M_p^2 X^3 X^{\prime} - m^2 M_p^2\sigma X -\frac{1}{2}m^2
 \sigma^2 X^\prime=0,
\end{equation}
which immediately gives
\begin{equation}
N_\sigma = \frac{X^\prime}{X} = \frac{2}{3}\frac{3(\frac{1}{2}m^2 \sigma^2 X^{-3})}{12 \Gamma^2 - \frac{1}{2}m^2 \sigma^2 X^{-3}}\frac{1}{\sigma} = \frac{2}{3}r\frac{1}{\sigma}.
\label{a3}
\end{equation}
The curvature perturbation is given by 
\begin{equation}
\zeta=N_\sigma \delta \sigma= \frac{2}{3}r\frac{\delta \sigma}{\sigma},
\label{b4}
\end{equation}
which is a standard result of curvaton.

In order to calculated $f_{NL}$, we simply have to make derivative of Eq.~(\ref{a3}) with respect with $\sigma$ once more and obtain
\begin{equation}
N_{\sigma\sigma}=-\frac{2}{3}r\frac{1}{\sigma^2}+\frac{2}{3}r_{\sigma} \frac{1}{\sigma}.
\label{b5}
\end{equation}
In order to calculate $r_\sigma$, we have to make derivative of $r$ with respect to $\sigma$ by using the last expression of Eq.~(\ref{eq5}) to obtain
\begin{equation}
r_\sigma= \frac{2r}{\sigma}-\frac{4}{3\sigma}r^2-\frac{2}{3 \sigma}r^3,
\label{r2}
\end{equation}
where Eqs.~(\ref{a3}) and (\ref{handy}) has been used.
Therefore we have
\begin{equation}
f_{NL}=\frac{5}{6}\frac{N_{\sigma\sigma}}{N_\sigma^2}=\frac{5}{4r}-\frac{5}{3}-\frac{5}{6}r.
\end{equation}
Similarly, in order to calculate $g_{NL}$, we can substitute Eqs.~(\ref{a3}) and (\ref{r2}) into Eq.~(\ref{b5}) to write
\begin{equation}
N_{\sigma\sigma}=\left( \frac{1}{\sigma}-\frac{4r}{3\sigma}-\frac{2r^2}{3\sigma} \right)N_\sigma,
\end{equation}
and then make derivative with respect to $\sigma$. Finally we can obtain
\begin{equation}
\frac{54}{25}g_{NL}=\frac{N_{\sigma\sigma\sigma}}{N_\sigma^3}=-\frac{9}{r}+\frac{1}{2}+10r+3r^2.
\end{equation}

\end{document}